\documentclass[12pt,a4paper]{article}
\usepackage{latexsym,graphicx,amssymb,amsmath}

\usepackage{setspace,bm}

\newcommand{\be}{\begin{equation}}
\newcommand{\ee}{\end{equation}}
\newcommand{\bea}{\begin{eqnarray}}
\newcommand{\eea}{\end{eqnarray}}

\newcommand{\rt}[1]{{}}

\makeatletter
\long\def\unmarkedfootnote#1{{\long\def\@makefntext##1{##1}\footnotetext{#1}}}
\makeatother

\begin{document}
{\allowdisplaybreaks

\title{Counterterm resummation for 2PI-approximation \\
in constant background}

\author{A. Patk{\'o}s$^{1a,b}$ \vspace*{0.05cm}\\
{\it $^a$Department of Atomic Physics, E{\"o}tv{\"o}s University}
\vspace*{0.05cm}
\\
{\it $^b$Research Group for Statistical and Biological Physics}\\
{\it of the Hungarian Academy of Sciences}\\
{\it H-1117 Budapest, Hungary}
\vspace*{0.15cm}\\
Zs. Sz{\'e}p$^{2}$\vspace*{0.05cm}\\
{\it Research Institute for Solid State Physics and Optics of the}\\
{\it Hungarian Academy of Sciences, H-1525 Budapest, Hungary}
}

\unmarkedfootnote{E-mail:
$^1$patkos@ludens.elte.hu, $^2$szepzs@achilles.elte.hu}

\date{}

\maketitle
\begin{abstract}
Explicit counterterm construction is presented in a symmetry breaking
background for the 2PI effective action of the self-interacting 
one-component scalar field including skeleton diagrams 
to ${\cal O}(\lambda^2)$. The applied strategy is an extension of our
treatment of the 2PI-Hartree approximation \cite{fejos08}.
The procedure is also applied to the $O(N)$ model.
\end{abstract}

\section{Introduction}

Renormalisation invariance is a powerful concept in constructing field
theoretical models describing phenomena in a restricted range of momenta. It
imposes insensitivity requirement for the physical predictions to the choice
of the maximal momentum if the range characterising the phenomenon under
investigation is much below that value. Renormalisable models represent a
specific subclass where the maximal momentum could be sent to infinity. In
view of triviality of scalar field theories, renormalisability just ensures
that high precision predictions calculated with finite (but large) cut-off
in such models will not depend sensitively on the cut-off.

Renormalisability criteria for perturbatively computed physical quantities
as power series of the renormalised couplings represent a classical piece of
knowledge \cite{collins84}. One constructs at every order also the
corresponding approximation of the counterterm Lagrangian, which cancels the
cut-off dependence of some loop integrals. The renormalisation of partially
resummed perturbative series is nontrivial because in association with the
selected subset of the perturbative diagrams one has to perform also the
resummation of appropriately chosen counterterm diagrams.

In the past 6-7 years considerable progress has been achieved in
constructing renormalisable two-particle-irreducible (2PI) approximate
solutions of perturbatively renormalisable theories. The 2PI approximation
was developed into a valuable non-perturbative tool in studies of collective
features of quantum field theories both in and out of equilibrium.  With its
help one can investigate the evolution of the spatial distribution of ground
state field condensates as well as propagators of their quasi-particle
excitations.

In a series of papers \cite{vhees02a,vhees02b,vhees02c} van Hees and Knoll
proved renormalisability of the 2PI approximate solutions in the real time
formulation of quantum field theories with help of the subtraction procedure
designed by Bogoliubov, Parasiuk, Hepp, and Zimmermann (BPHZ). Their study
focused on the symmetric phase with no condensate in the ground state. This
line was continued in 2003-04 by Blaizot, Iancu and Reinosa
\cite{blaizot03,blaizot04} having demonstrated iteratively that
self-consistently solved propagator equations resum infinite subsets of
different orders of the perturbation theory. Mainly based on diagrammatic
analysis of the iterations they also constructed the counterterms necessary
to balance the cut-off dependence of the loop integrals appearing in the
self-consistent equations. Moreover they pointed out the importance of the
large momentum asymptotics of the propagator for the determination of the
cut-off dependence of the counter couplings. Related investigations were
done also using the auxiliary field formulation of $\Phi^4$ theories (e.g.
after Hubbard-Stratonovich transformation)
\cite{cooper04,cooper05,calzetta87} and also concentrating on the problem of
gauge fixing parameter dependence of 2PI approximation in gauge theories
\cite{arrizabalaga02,carrington05}.

A complete description of how to renormalise arbitrary $n$-point functions
derived in 2PI-formalism was presented by Berges, Bors\'anyi, Reinosa and
Serreau in Ref.~\cite{berges05a}, also for non-vanishing expectation value
of the scalar field. The detailed analysis of that paper was mainly based
also on diagrammatics and explicit counterterm formulae were displayed only
partially. Since then several numerical implementations were reported for
scalar theories \cite{arrizabalaga05,arrizabalaga06, berges05b} and also for
QED \cite{reinosa06,borsanyi08}. The symmetry aspects of the 2PI-formalism
was investigated in \cite{vhees02c}. It was shown that only the so-called
``external N-point functions'', {\it i.e.} derivatives of the effective
action with respect to the mean field fulfill Ward identities. In
particular, the self-consistent propagator differs from the external
two-point function and in consequence it violates Goldstone's theorem. The
validity of Ward identities was discussed recently in the context of QED in
\cite{reinosa07}.

Investigations of the renormalisation were performed also for a related
resummation scheme using local (momentum-independent) insertions, the
so-called two-particle-point-irreducible (2PPI) approximations. The
renormalisation of this scheme was proved using counterterms
\cite{verschelde01} and the scheme was used in numerical works both in
equilibrium \cite{verschelde02} and out of equilibrium
\cite{baacke03,baacke04}. Recently in Ref.~\cite{toni07} a renormalisation
group inspired approach was developed to investigate the connection between
momentum-dependent resummation and renormalisation. There a comparison of
the method with renormalisation methods relying on Bethe-Salpeter-like
equations was also given.

The purpose of the present paper is to deduce the resummed counterterms in a
transparent way, and provide explicit formulae as far as possible for the
counter couplings. In a previous paper \cite{fejos08} we have shown for the
2PI-Hartree approximation that results of the iterative renormalisation of
Ref.~\cite{blaizot04} can be obtained also with help of a one-step
substitution. In this paper we demonstrate first the applicability of our
procedure to all cases where the large momentum asymptotics of the
propagator remains the same as it was in the tree level approximation. In
particular, we find that in such cases there is no need to use any auxiliary
Bethe-Salpeter-like equation for the determination of the 4-point
counter-couplings. This feature can be studied neatly on the example of the
complete two-loop skeleton diagram set of the 2PI effective action in a
constant, homogeneous symmetry breaking background. The basic definitions
and notations of the two models considered in this paper are introduced in
Section~2. In Section~3 the construction is performed first for the case of
the one-component real field. It is applied to the more involved case of the
$O(N)$ model in Section~5. The presentation will be detailed and still more
compact than was the case of previous analyses.

The generalisation of our procedure to the case of the modified large
momentum behaviour of the propagator is discussed for the one-component
scalar field in Section~4. It leads to a new important intermediate step in
designing the algorithm of the counterterm determination. Namely, the
separation of the divergent pieces of different skeleton diagrams has to be
based now on an auxiliary propagator whose asymptotics coincides with the
exact large momentum behaviour and is independent of the possibly
environment dependent infrared data (masses, condensates) of the theory
\cite{blaizot04}. The exact propagator is expanded around this propagator,
which eventually determines the counterterms. This analysis becomes
unavoidable when also the so-called ``basket-ball'' diagram is added to the
set of skeleton diagrams determining the 2PI effective action. It is in
Section~6 where we extend the analysis to the $O(N)$ model. A central
question of these sections will be, how the form of subdivergences changes,
whether all subdivergences can be expressed as products of a unique
(possibly environment dependent) function and some explicit cut-off
dependent factors, or one has to consider several independent ``mixed''
products. It turns out that the relevant contribution from the
``basket-ball'' diagram can be decomposed into the sum of terms from which
some are proportional to powers of the background field, some to the finite
part of the tadpole integral, therefore no new type of subdivergence shows
up. This decomposition is determined by an integral kernel solving a
Bethe-Salpeter type equation, which introduces quite naturally a resummation
for certain 4-point functions.

As a consequence, the classification of the divergent pieces remains
unchanged relative to the classes found already in the 2PI-Hartree
approximation \cite{fejos08}: one finds overall divergences proportional to
the zeroth and second power of the background field and subdivergences
proportional to the finite part of the independent tadpole integrals. Their
coefficients are required to vanish separately. In general, the number of
renormalisation conditions arising from this requirement is larger than the
number of independent counter couplings. Renormalisability of the 2PI
approximation means beyond the determination of all resummed counter
couplings also the demonstration that these supplementary equations do not
contradict those which are used for extracting the counterterms.

The summary of our findings is given in Section~7 where we shortly outline
possible approaches to renormalised (cut-off independent) numerical
calculations. Some technical details relevant to the discussion presented in
the main text are given in two Appendices.

\section{The models and some notations
\label{sec:model}}   
  
The complete static ${\cal O}(\lambda^2)$ skeleton 2PI functional
in the broken phase of the real one-component $\Phi^4$ model, 
characterised by the vacuum expectation value
$\Phi=v$ is of the following form \cite{arrizabalaga05}:
\bea   
V[v,G]&=&\frac{1}{2}m^2v^2+\frac{\lambda}{24}v^4
-\frac{i}{2}\int_p\big[\ln G^{-1}(p)+D^{-1}(p)G(p)\big]
+\frac{\lambda}{8}\left(\int_p G(p)\right)^2\nonumber\\   
&-&\frac{i\lambda^2}{12}v^2\int_k\int_p G(p)G(k)G(p-k)
-\frac{i\lambda^2}{48}\int_k\int_p\int_q G(q)G(p)G(k)G(p+k+q)
+V_{ct}[v,G],
\label{lambda2-2pi-action}
\eea
where  $i D^{-1}(p)=p^2-m^2-\lambda v^2/2$ is the tree-level inverse 
propagator. The counterterm functional is given by
\be   
V_{ct}[v,G]=\frac{1}{2}\delta m^2_0 v^2+\frac{\delta\lambda_4}{24}v^4+   
\frac{1}{2}\left(\delta m_2^2
+\frac{\delta\lambda_2}{2}v^2\right)\int_p G(p)
-\frac{\delta Z}{2}\int_p p^2 G(p)
+\frac{\delta\lambda_0}{8}\left(\int_p G(p)\right)^2.
\label{lambda2-2pi-counter}   
\ee   
Note, that there is no counterterm diagram involving the
``setting-sun'' or the ``basket-ball'' diagrams. Such diagrams would
be necessary only when going beyond the present level of the 2PI
truncation, as will become clear in Section 4.

The generalisation of the effective action to the $N$-component
scalar field with $O(N)$ symmetry in the symmetry breaking background
is expressed with the following representation of the 
tree-level and exact propagators:
\be
i D_{ab}^{-1}(k)=\left[k^2-m^2-\frac{\lambda v_c^2}{6N}\right]
(P_{ab}^\sigma+P_{ab}^\pi)-\frac{\lambda v_c^2}{3N} P_{ab}^\sigma,
\qquad
G_{ab}(k)=G_\sigma(k)P_{ab}^\sigma+G_\pi(k)P_{ab}^\pi.
\label{prop_decomp}
\ee
Here $P_{ab}^\sigma=\hat v_a\hat v_b$, 
$P_{ab}^\pi=\delta_{ab}-\hat v_a\hat v_b$ are the relevant orthogonal
projectors. $\hat v^a,$ $a=1\dots N$ is the component of the unit vector
giving the orientation of the symmetry breaking constant background.

In terms of the two propagating modes the 2PI-action has the form:
\bea
V[v_a,G_\sigma,G_\pi]&=&\frac{1}{2}m^2v_a^2+\frac{\lambda}{24N}(v_a^2)^2
\nonumber\\
&-&
\frac{i}{2}\int_k\big[(N-1)\ln G_\pi^{-1}(k)+\ln G_\sigma^{-1}(k)+
(N-1)D_{\pi}^{-1}(k)G_\pi(k)+D_{\sigma}^{-1}(k)G_\sigma(k)\big]
\nonumber\\
&+&\frac{\lambda}{24N}\left[
3\left(\int_k G_\sigma(k)\right)^2+(N^2-1) \left(\int_k G_\pi(k)\right)^2
+2(N-1)\int_k G_\pi(k) \int_p G_\sigma(p) \right]
\nonumber\\
&-&\frac{i\lambda^2}{36N^2}v_a^2\int_k\int_p\Big[
3G_\sigma(k) G_\sigma(p)+(N-1)G_\pi(k)G_\pi(p)\Big] G_\sigma(p+k)
\nonumber\\
&-&\frac{i\lambda^2}{144N^2}\int_k\int_p\int_q
\Big[ (N-1) \Big( (N+1) G_\pi(k)G_\pi(p)
+2 G_\sigma(k)G_\sigma(p)\Big)
G_\pi(q)G_\pi(k+p+q)\nonumber\\
&&\qquad\qquad\qquad\quad
+3G_\sigma(k)G_\sigma(p)G_\sigma(q)G_\sigma(k+p+q)\Big]+V_{ct},
\label{lambda2-2pi-action-ON}
\eea
where $i D_{\pi}^{-1}(p)=p^2-m^2-\lambda v_a^2/(6 N), 
i D_{\sigma}^{-1}=p^2-m^2-\lambda v_a^2/(2 N)$ are the tree-level 
pion and sigma inverse propagators. In this case the counterterm 
functional reads
\bea
V_{ct}&=&\frac{1}{2}\delta m_0^2v_a^2+\frac{\delta\lambda_4}{24N}(v_a^2)^2
+\frac{1}{2}\left(\delta m_2^2+\frac{\delta\lambda_2^A}{6N}v^2
\right)\int_k \big(G_\sigma(k)+(N-1)G_\pi(k)\big)+
\frac{\delta\lambda_2^B v^2}{6N}\int_k G_\sigma(k)
\nonumber\\
&+&\frac{1}{24N}\left\{\delta\lambda_0^A\left[
\int_k\big((N-1)G_\pi(k)+G_\sigma(k)\big)\right]^2
\!\!+2\delta\lambda_0^B\Bigg[(N-1)\left[
\int_k G_\pi(k)\right]^2+\left[\int_k G_\sigma(k)\right]^2\Bigg]\right\}
\nonumber
\\
&-&\frac{\delta Z}{2}\int_p p^2 G_\sigma(p)
-(N-1) \frac{\delta Z}{2} \int_p p^2 G_\pi(p).
\label{lambda2-2pi-counter-ON}
\eea 
The introduction of the two quartic counterterms $\delta\lambda_0^A$
and $\delta\lambda_0^B$ corresponds to the two independent $O(N)$-invariant
combinations $G_{ab}G_{ba}$ and $G_{aa}G_{bb}$. 
Similarly $\delta m_2^2,$ $\delta\lambda_2^A,$ and $\delta\lambda_2^B$ 
are introduced to cancel divergences emerging from the
 two terms, $\delta_{ab} G_{ab}$ and $\hat v_a\hat
v_b G_{ab}$, appearing in $D^{-1}_{ab} G_{ba}$. 

In the one-component model divergent contributions to 
the equation of state $\delta V/\delta v=0$
and to the propagator equation $\delta V/\delta G(k)=0$ 
will be expressed with the divergent pieces of the
following integrals:
\bea
T[G]&:=& \int_k G(k)=T_\textnormal{div}+T_F[G],
\label{Eq:tadpole}\\
I(p,G)&:=& -i\int_k G(k)G(k+p)=I_\textnormal{div}+I_F(p,G),
\label{Eq:bubble}\\
S(p,G)&:=& -i\int_k\int_q G(k)G(q)G(k+q+p)=S_\textnormal{div}(p,G)+S_F(p,G).
\label{Eq:setting-sun}
\eea
For the $N$-component case one introduces correspondingly $T_\alpha,
I_{\alpha\beta}, S_{\alpha\beta\gamma}$, where the indices can take the
values $\alpha=\{\pi,\sigma\}$. Separation of the 
divergent pieces is explicitly presented in the Appendices. 
Without introducing any particular notation, the finite parts of these
integrals would include temperature and density dependent contributions.

The non-perturbative construction of the counterterms will be presented in
two stages. First, we restrict the discussion to the two-loop truncation of
the effective actions given in (\ref{lambda2-2pi-action}) and
(\ref{lambda2-2pi-action-ON}). In this case the asymptotic behaviour of the
propagator remains as it is at tree-level. In a second step we shall discuss
the effect of adding the ``basket-ball'' diagram, which changes the large
momentum asymptotics. We believe that the structured presentation makes more
clear the algorithm underlying the somewhat complex formulae and facilitates
its prospective numerical implementations.

\section{Analysis of the two-loop truncation: one component 
real scalar  field}

In this truncation one omits the contribution of the last but one term of
(\ref{lambda2-2pi-action}). The equation for the propagator reads as
\be
i G^{-1}(p)=p^2-m^2-\delta m_2^2-\frac{1}{2}(\lambda+\delta\lambda_2)v^2
-\frac{1}{2}(\lambda+\delta\lambda_0)T[G]-\frac{1}{2}\lambda^2v^2I(p,G).
\label{prop-eq}
\ee
The large $p$ asymptotics of the propagator is unchanged compared to
the tree-level one, and we parametrise the finite self-energy of the
exact propagator as
\be
i G^{-1}(p)=p^2-M^2-\Pi(p).
\label{eq:G}
\ee
The splitting of the finite self-energy into the two pieces
\be 
M^2=m^2+\frac{\lambda}{2}v^2+\frac{\lambda}{2}T_F[G],\qquad
\Pi(p)=\frac{\lambda^2}{2}v^2I_F(p,G), \qquad
\lim_{p\rightarrow\infty}\frac{\Pi(p)}{p^2}\rightarrow 0,
\label{Eq:finite_self-energy}
\ee
separates the momentum-dependent part of the finite self-energy ($\Pi(p)$),
coming from the setting-sun integral of the effective-action 
(\ref{lambda2-2pi-action}) and the momentum-independent contribution 
($M^2$) which comes from the other two-loop piece of $V[v,G]$. 
Although this latter term formally corresponds to the 2PI-Hartree
truncation discussed in our previous paper \cite{fejos08}, its
self-consistent determination takes into account the effect of $\Pi(p).$

The divergence cancellation condition which determines the
counterterms $\delta m_2^2,\delta\lambda_2,$ and $\delta\lambda_0$ is
obtained by subtracting (\ref{eq:G}) with the parametrisation
(\ref{Eq:finite_self-energy}) from (\ref{prop-eq}) and has the form:
\be
0=\delta m_2^2+\frac{1}{2}\delta\lambda_2v^2+
\frac{1}{2}(\lambda+\delta\lambda_0)T_\textnormal{div}+
\frac{1}{2}\delta\lambda_0 T_F[G]+
\frac{1}{2}\lambda^2v^2I_\textnormal{div}.
\label{prop-div}
\ee

The equation of state reads as
\be
v\left(m^2+\delta m^2_0+\frac{1}{6}(\lambda+\delta\lambda_4)v^2+
\frac{1}{2} (\lambda+\delta\lambda_2)T[G]+
\frac{1}{6}\lambda^2S(0,G)\right)=0.
\label{state-eq}
\ee 
The cancellation condition for the divergent pieces of this equation amounts
to
\be
0=\delta m^2_0+\frac{1}{6}\delta\lambda_4v^2+\frac{1}{2}(\lambda+
\delta\lambda_2)T_\textnormal{div}+
\frac{1}{2}\delta\lambda_2T_F[G]+\frac{1}{6}\lambda^2
S_\textnormal{div}(0,G).
\label{state-div}
\ee
From this equation one should determine $\delta m^2_0,
\delta\lambda_4$ making use of the previously determined
expression for $\delta\lambda_2$.

The regularised expressions of the counter couplings as determined
from (\ref{prop-div}) and (\ref{state-div}) give regularisation independent
solutions when used in the regularised equations (\ref{prop-eq}) and
(\ref{state-eq}). In our discussion we shall have in mind cut-off 
regularisation of the divergent integrals. Since the divergent part of
these integrals is separated using propagators which are not
sensitive to the infrared features of the theory, their entire zero
temperature part will be subtracted.
 
With help of Appendix A one finds the following expressions for the
divergent pieces of the basic integrals:
\bea
T_\textnormal{div}&=&T_d^{(2)}+(M^2-M_0^2)T_d^{(0)}+
\frac{1}{2}\lambda^2v^2 T_d^{(I)},\qquad\qquad\qquad
I_\textnormal{div}=T_d^{(0)},\nonumber\\
S_\textnormal{div}(0,G)&=&S_{P V}(0)+3T_d^{(0)}T_F+3(M^2-M_0^2)
\left((T_d^{(0)})^2+T_d^{(I)}\right)
+\frac{3}{2}\lambda^2v^2\left(T_d^{(I)}T_d^{(0)}+T_d^{(I,2)}\right),
\label{TIS-div}
\eea
The cut-off dependent integrals 
$S_{P V}(0), T_d^{(2)}, T_d^{(0)}, T_d^{(I)},$ and $T_d^{(I,2)}$ 
are defined in Appendix A. $M_0^2$
is a free parameter introduced in the auxiliary propagator
(\ref{eq:aux-prop}) to ensure the infrared finiteness of all
integrals. Exploiting the definition of $M^2$ given in
(\ref{Eq:finite_self-energy}) when using (\ref{TIS-div}) in
(\ref{prop-div}) and (\ref{state-div}), one separates the coefficients
of $v^0, v^2$ and of $T_F[G]$ exactly as it was done when we discussed
the renormalisation of the 2PI-Hartree approximation
\cite{fejos08}. The following six equations arise:
\bea
0=&&\delta m_2^2+\frac{1}{2}(\lambda+\delta\lambda_0)
\left[T_d^{(2)}+(m^2-M_0^2)
T_d^{(0)}\right],\nonumber\\
0=&&\delta\lambda_2+\frac{1}{2}\lambda(\lambda+\delta\lambda_0)
\left(T_d^{(0)}+\lambda T_d^{(I)}\right)+\lambda^2T_d^{(0)},\nonumber\\
0=&&\delta\lambda_0+\frac{1}{2}\lambda(\lambda+\delta\lambda_0)T_d^{(0)},
\nonumber\\
0=&&\delta m^2_0+\frac{1}{2}(\lambda+\delta\lambda_2)
\left[T_d^{(2)}+(m^2-M_0^2) T_d^{(0)}\right]+\frac{1}{2}\lambda^2(m^2-M_0^2)
\left((T_d^{(0)})^2+T_d^{(I)}\right)+
\frac{1}{6}\lambda^2S_{P V}(0),\nonumber\\
0=&&\delta\lambda_4+\frac{3}{2}\lambda(\lambda+\delta\lambda_2+\lambda^2
T_d^{(0)})\left(T_d^{(0)}+\lambda T_d^{(I)}\right)+\frac{3}{2}\lambda^3
\left(T_d^{(I)}+\lambda T_d^{(I,2)}\right),\nonumber\\
0=&&\delta\lambda_2+\frac{1}{2}\lambda(3\lambda+\delta\lambda_2)T_d^{(0)}+
\frac{1}{2}\lambda^3\left((T_d^{(0)})^2+T_d^{(I)}\right),
\label{Eq:1comp_counterterms}
\eea
where the first three equations come from (\ref{prop-div}), while the
last three from (\ref{state-div}). The expression for
$\delta\lambda_0$ determined from the third equation is identical to
the expression which we would have obtained in 2PI-Hartree approximation.
If one determines from the first five equations $\delta m^2_0,\delta
m_2^2, \delta\lambda_0,\delta\lambda_2,$ and $\delta\lambda_4$, then
the sixth serves for a consistency check of the renormalisation. It
can be verified by direct substitution of the previously obtained
counter coupling expressions. The reason behind this consistency is 
that when the exact propagator is expanded around the auxiliary propagator 
the same integral appears both in the momentum independent part of the 
self-energy and in the contribution of the setting-sun to the equation of 
state. In the former case the divergent part of this integral denoted with
$T_d^{(I)}$ appears as an  overall divergence proportional to $v^2$, 
which can be seen by looking at the  last term of $T_\textnormal{div}$ in
(\ref{TIS-div}). In the second case it appears also as a
subdivergence, that is proportional to $T_F[G]$ (see the third term 
of $S_\textnormal{div}$ in (\ref{TIS-div})). This structural correspondence 
is illustrated in Fig.~\ref{Fig:1comp}.

We remark that the results in (\ref{Eq:1comp_counterterms}) obtained with
the present method reproduce results announced by Arrizabalaga and Reinosa
in a conference poster \cite{arrizabalaga06} (explicit expressions for the
counterterms were given only on the poster, but not in the Proceedings). 
The use of these expressions in (\ref{prop-eq}) and (\ref{state-eq}) makes
these equations renormalised (cut-off independent).

\begin{figure}
\centerline{
\includegraphics[keepaspectratio,width=0.3\textwidth,angle=0]{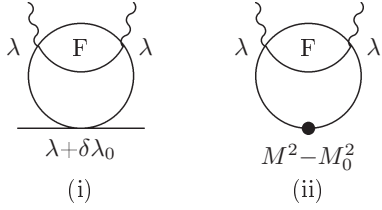}}
\caption{
Illustration of the appearance of the same radiative structure
producing (i) the divergence denoted with $T_d^{(I)}$ in the equation for
the inverse propagator and (ii) the subdivergence of the equation of
state. Internal lines represent the auxiliary propagators $G_{PV}.$
Wiggly line represents the background $v.$ Letter `F' refers to the
finite bubble contribution $I_{PV,F}.$}
\label{Fig:1comp}
\end{figure}

\section{Adding the ``basket-ball'' diagram}

Keeping all terms in the ${\cal O}(\lambda^2)$ truncation 
of the skeleton decomposition of the 2PI potential the finite 
self-energy contains three terms:
\bea
&
i G^{-1}(p)=p^2-\Sigma(p),\qquad \Sigma(p)=M^2+\Pi_0(p)+\Pi_2(p),\nonumber\\
&\displaystyle
M^2=m^2+\frac{1}{2}\lambda v^2
+\frac{\lambda}{2}T_F[G],\qquad \Pi_0(p)=\frac{1}{2}\lambda^2v^2I_F(p),
\qquad \Pi_2(p)=\frac{\lambda^2}{6}S_F(p).
\label{eq:G-a}
\eea
The cancellation of the divergences of the propagator equation
amounts to the equality
\be
0=\delta m^2_2+\frac{1}{2}\delta\lambda_2v^2+\frac{1}{2}
(\lambda+\delta\lambda_0)T_\textnormal{div}+
\frac{1}{2}\delta\lambda_0 T_F[G]+
\frac{1}{2}\lambda^2v^2I_\textnormal{div}+
\frac{1}{6}\lambda^2S_\textnormal{div}(p)-p^2\delta Z.
\label{prop-div-a}
\ee

The new feature of the analysis is the fact that the large momentum
behaviour of the propagator changes due to the setting-sun contribution. 
The new leading asymptotics is accompanied also by an infinite wave function
renormalisation. The situation is complicated by the fact that the
asymptotics can be determined only in the course of the numerical solution
of the renormalised propagator equation (the equation of state has no
influence on the asymptotics). Nevertheless, one proceeds by treating
separately the leading asymptotics not depending on any infrared parameter
of the theory:
\be
\Pi_2(p)=\Pi_a(p)+\Pi_{2,0}(p)+\Pi_r(p), \qquad \lim_{p\rightarrow\infty}
\frac{\Pi_a(p)}{p^2}\sim (\ln p)^\beta,\qquad \lim_{p\rightarrow \infty}
\frac{\Pi_{2,0}(p)}{p^2}\rightarrow 0,\qquad \Pi_r(p)\sim p^{-2}.
\label{Eq:Pi2_decomp}
\ee
The splitting is somewhat arbitrary and this freedom can be exploited 
to find simple-to-handle expressions for the counter-couplings.
There is a corresponding splitting of the setting-sun contribution:
\be
S_F(p)=S_{a,F}(p)+S_{0,F}(p)+S_r(p),
\ee
where $S_a(p)$ is defined as the setting-sun diagram computed with 
a conveniently introduced, infrared stable auxiliary propagator
which asymptotically coincides with the exact propagator:
\be
i G_a^{-1}(p)=p^2-M_0^2-\Pi_a(p),\qquad G(p)=G_a(p)+\delta G(p),\qquad
\delta G(p)\sim {\cal O}(p^{-4}).
\label{Eq:Ga_def}
\ee 
Weinberg's theorem \cite{weinberg59} guarantees that the 
self-consistent solution of this equation  for the self-energy
(inverse propagator) behaves asymptotically as $\sim p^2 (\ln p)^\beta,$
where $\beta$ can be determined only numerically. 

Having introduced all the relevant notations we proceed now with the
analysis of the divergences. The momentum dependent divergent 
part of $S_a(p)$ is removed by the wave function renormalisation 
counterterm:
\be
\delta Z=\frac{\lambda^2}{6} \frac{\partial}{\partial p^2} 
S_a^\textnormal{div}(p).
\ee

Again, the renormalisation of the bubble integral is the simplest:
\be
I(p)=I_\textnormal{div}+I_F(p),\qquad 
I_\textnormal{div}:= T_a^{(0)},\qquad
I_F(p)=-i\int_k\big[G(k)G(k+p)-G_a^2(k)\big].
\label{Eq:Idiv}
\ee 
The tadpole integral is analysed with help of the identity given in
(\ref{Eq:azonossag2_G})
\be
T[G]=T_a^{(2)}+(M^2-M_0^2)T_a^{(0)}+\frac{1}{2}\lambda^2v^2T_a^{(I)}-
i\int_k G_a^2(k)\Pi_{2,0}(k)+T_F^{(1)}[G].
\label{tadpole-a}
\ee
The finite contribution denoted with $T_F^{(1)}[G]$ consists of two pieces.
The first is the contribution of $G_r(p)$ and the second one 
is the contribution of 
$I_F(k)-I_{a,F}(k),$ which arises when in $\Pi_0(k)$ one replaces 
$I_F(k)$ by $I_{a,F}(k)$ to obtain the divergence $T_a^{(I)}.$
One finds the definitions of $T_a^{(2)}, T_a^{(0)}$ and of $T_a^{(I)}$ 
in Appendix B.

The determination of $\Pi_{2,0}(k)$ is needed to understand the
contribution of the last explicit integral term in the expression of
$T[G]$ given in (\ref{tadpole-a}). It implies
 the divergence analysis of $S(p)$.  With help of the 
decomposition $G(p)=G_{a}(p)+\delta G(p)$ (see eq.~(\ref{Eq:azonossag2_G}))
one can disentangle the pieces of the
setting-sun integral defined in (\ref{Eq:setting-sun}), which
potentially contain divergences:
\be
S(p)=S_a(p)+3\int_k\delta G(k)I_a(k+p)+S_r^{(1)}(p), 
\qquad S_r^{(1)}(p)\sim p^{-2}.   
\label{eq:S-a}
\ee
The momentum dependent part of the first term determines the wave function
renormalisation, as we have already indicated, and also $\Pi_a(p)$. 
The last term gives a contribution to $\Pi_r(p).$ 

One identifies $S_{0,F}(p)$ with the contribution of the integral in 
(\ref{eq:S-a}) which is finite and does not vanish asymptotically for 
large $p$. Using the expression of $\delta G$ given in 
(\ref{Eq:azonossag2_G}), simple power counting shows that its $G_r(k)$ term  
gives a second contribution to $\Pi_r(p).$ Moreover, in the expression 
of $\delta G(k)$ one can replace $I_F(k)$ by $I_{a,F}(k)$, 
since the difference gives a third contribution to $\Pi_r(p).$ 
Taking in (\ref{eq:S-a}) the finite part of the asymptotic bubble $I_a$, 
one observes that the integral has a logarithmic divergence which is captured 
by changing $I_{a,F}(k+p)$ to $I_{a,F}(k).$ It is their difference
which 
contributes 
to $S_{0,F}(p).$ To improve the convergence of the integral in the
ultraviolet, one exploits the symmetry property of the integrand and uses 
instead of $I_{a,F}(k+p)-I_{a,F}(k)$ the kernel
\be
K(p,k)=\frac{\lambda^2}{2}\Big[I_{a,F}(k+p)+I_{a,F}(k-p)-2I_{a,F}(k)\Big],
\label{Eq:BS_kernel}
\ee
which is symmetric under the reflection of  $k\rightarrow -k.$ Note,
that the kernel is not symmetric upon the interchange of $k$ and $p,$
that is $K(p,k)\ne K(k,p).$ This is to be contrasted with the
symmetric kernel of Ref.~\cite{blaizot04}, denoted with $\Lambda(p,k)$
which, however, is not finite. 

With the above steps the following inhomogeneous 
integral representation is obtained for $S_{0,F}(p)$: 
\be
\frac{\lambda^2}{6}S_{0,F}(p)=-\frac{i}{2}
\int_k G_a^2(k) K(p,k)\left[M^2-M_0^2+
\frac{\lambda^2}{2} v^2I_{a,F}(k)+\frac{\lambda^2}{6}S_{0,F}(k)\right].
\label{eq:S0F-finite}
\ee
This can be ``inverted'' iteratively to take the form
\be
\frac{\lambda^2}{6}S_{0,F}(p)=\frac{1}{2}
(M^2-M_0^2)\left(-i\int_k\Gamma(p,k)G_a^2(k)\right)+
\frac{1}{4}\lambda^2 v^2\left(-i\int_k\Gamma(p,k)G_a^2(k)I_{a,F}(k)\right),
\label{Eq:S0F_sol}
\ee
where the kernel $\Gamma(k,p)$ fulfills the Bethe-Salpeter-like equation
\be
\Gamma(p,k)=K(p,k)-\frac{i}{2}\int_q G_a^2(q)K(p,q)\Gamma(q,k).
\label{Eq:BS}
\ee
\indent The virtue of the representation in (\ref{Eq:S0F_sol}) is
that it decomposes $S_{0,F}(p)$ into the sum of terms proportional to
$v^0, v^2$ and $T_F[G]$. It is not a solution for $S_{0,F}$, 
since $T_F[G]$ contains it implicitly, but it means that no new 
independent finite quantity appears in $\Pi_{2,0}(p)$ relative 
to the analysis in Section~3. In other words, the subdivergences 
at this level are found again by looking at the coefficient of 
$T_F[G]$ in the divergence produced by 
$\Pi_{2,0}(k)=\lambda^2 S_{0,F}(p)/6$ 
through the integral in (\ref{tadpole-a}).

Using the expression of $M^2$ from (\ref{eq:G-a}) in  (\ref{Eq:S0F_sol})
one obtains
\be
\Pi_{2,0}(p)=\frac{1}{2}(m^2-M_0^2)\Gamma_0(p)+\frac{v^2}{4}\left[\lambda
\Gamma_0(p)+\lambda^2\Gamma_1(p)\right]+\frac{\lambda}{4}T_F[G]\Gamma_0(p),
\label{eq:S0F}
\ee
where for the finite integrals we introduced the following
shorthand notations:
\be
\Gamma_0(p)=-i\int_k\Gamma(p,k)G_a^2(k),\qquad 
\Gamma_1(p)=-i\int_k \Gamma(p,k)G_a^2(k)I_{a,F}(k).
\ee 
The coefficients $\Gamma_0(p)$ and $\Gamma_1(p)$  multiplying the quantities
sensitive to the infrared parameters (e.g. $m^2-M_0^2, v^2, T_F[G]$) are fully 
determined by $G_a(p)$ and have logarithmic large-$p$ asymptotics!

With the expression of $\Pi_{2,0}(p)$ given in (\ref{eq:S0F})
one returns to the expression of the tadpole integral
(\ref{tadpole-a}). Introducing for the divergences of the 
remaining integral term the following notations:
\be
D_0=-i\int_k G_a^2(k)\Gamma_0(k)\Big|_\textnormal{div},\qquad
D_1=-i\int_k G_a^2(k)\Gamma_1(k)\Big|_\textnormal{div},
\label{Eq:D01_def}
\ee
one obtains
\bea
T[G]&=&T_a^{(2)}+(m^2-M_0^2) \left(T_a^{(0)}+\frac{1}{2} D_0\right) 
+\frac{1}{2}\lambda^2 v^2 \left[
T_a^{(0)}+\lambda T_a^{(I)}+\frac{1}{2}
\big(D_0+\lambda D_1\big)\right]
\nonumber
\\
&&+\left[1+\frac{\lambda}{2}\left(T_a^{(0)}+\frac{1}{2} D_0\right)\right]
T_F[G].
\label{Eq:Tad_div_BS}
\eea
Note, that  the finite parts of the integrals given in 
(\ref{Eq:D01_def}) are added to $T_F^{(1)}[G]$ of (\ref{tadpole-a}) to give 
$T_F[G].$ The analysis of the divergences coming from the tadpole integral is
now complete.

In order to find the expressions of the counter couplings one
 needs also $S_\textnormal{div}$. Using 
(\ref{Eq:azonossag2_G}) in (\ref{eq:S-a}) one readily obtains:
\bea
\nonumber
S_\textnormal{div}(p)-\frac{6}{\lambda^2}p^2\delta Z
&=&S_{a,\textnormal{div}}(0)+3T_a^{(0)}(T[G]-T_a^{(2)})\\
&&-3i\left\{\int_k G_a^2(k)I_{a,F}(k+p)\left[M^2-M_0^2+
\frac{1}{2}\lambda^2 v^2I_{F}(k)+\Pi_{2,0}(k)\right]\right\}
\bigg|_\textnormal{div}
\nonumber\\
&=&S_{a,\textnormal{div}}(0)+3T_a^{(0)}(T[G]-T_a^{(2)})
+3\big(M^2-M_0^2\big) T_a^{(I)}
\nonumber
\\
&&+\frac{3}{2}\lambda^2v^2T_a^{(I,2)}
-3i\int_k G_a^2(k)I_{a,F}(k)\Pi_{2,0}(k)\Big|_\textnormal{div},
\label{Eq:Sdiv_final}
\eea
where $T[G]$ is given in (\ref{Eq:Tad_div_BS}) and the divergences of the
last term are expressible using (\ref{eq:S0F}).  In going from the
first to the second form of (\ref{Eq:Sdiv_final}), we are allowed to
replace both $I_F(k)$ and $I_{a,F}(k+p)$ by $I_{a,F}(k)$
because the differences give finite contributions.  It is remarkable
that again the infrared quantities ($m^2, M_0^2, v^2, T_F[G]$) appear
linearly with coefficients fully determined by $G_a$. Therefore the
zero temperature part of the integrals determining these coefficients
will be fully included into the counterterms!

Having the expressions of $I_\textnormal{div},$ $T_\textnormal{div},$ and
$S_\textnormal{div}$ (see (\ref{Eq:Idiv}),
(\ref{Eq:Tad_div_BS}), and (\ref{Eq:Sdiv_final})),  
one substitutes into the last one the definition of $M^2$ from
(\ref{eq:G-a}) and one proceeds to the decomposition of the divergence
cancellation condition for the propagator equation into the three
separate equations corresponding to the pieces
proportional to $v^0,$ $v^2,$ and $T_F[G],$ respectively:
\bea
0&=&\delta m_2^2+\frac{1}{2}(\lambda+\delta\lambda_0)
\left[T_a^{(2)}+(m^2-M_0^2)T_a^{(0)}\right]+
\frac{\lambda^2}{6}S_{a,\textnormal{div}}(0)
\nonumber\\
&&+(m^2-M_0^2)\left\{\frac{\lambda^2}{2}\left[
(T_a^{(0)})^2+T_a^{(I)}+\frac{1}{2}\tilde D_1\right]+
\frac{1}{4}\big(\lambda+\delta\lambda_0+\lambda^2T_a^{(0)}\big)D_0\right\},
\label{dm2-a}
\\
0&=&\delta\lambda_2+\frac{1}{2}\lambda
(\lambda+\delta\lambda_0)
\big(T_a^{(0)}+\lambda T_a^{(I)}\big)+\lambda^2T_a^{(0)}
\nonumber\\ 
&&+\frac{1}{2}\lambda^3\Big[(T_a^{(0)})^2+T_a^{(I)}+
\lambda\left(T_a^{(0)}T_a^{(I)}
+T_a^{(I,2)}\right)\Big]\nonumber\\
&&
+\frac{1}{4}\lambda\Big[\left(\lambda+\delta\lambda_0+\lambda^2T_a^{(0)}\right)
(D_0+\lambda D_1)+\lambda^2(\tilde D_1+\lambda D_2)\Big],
\label{dl2-a}
\\
0&=&\delta\lambda_0+\frac{1}{2}\lambda(\lambda+
\delta\lambda_0)T_a^{(0)}+\lambda^2\left[T_a^{(0)}+\frac{1}{2}\lambda
\left((T_a^{(0)})^2+T_a^{(I)}\right)\right]
\nonumber\\
&&+\frac{1}{4}\lambda\Big[\left(\lambda+\delta\lambda_0+
\lambda^2T_a^{(0)}\right)D_0+\lambda^2 \tilde D_1\Big].
\label{dl0-a}
\eea
The new quantities appearing in these equations are the following:
\bea
\tilde D_1=-i\int_k G_a^2(k) I_{a,F}(k) \Gamma_0(k)\Big|_\textnormal{div},
\qquad
D_2=-i\int_k G_a^2(k)I_{a,F}(k)\Gamma_1(k)\Big|_\textnormal{div}.
\label{Eq:D1T2_def}
\eea
Note, that $D_1\ne \tilde D_1$ because the kernel $K(p,k)$ 
given in (\ref{Eq:BS_kernel}) is asymmetric in its arguments. 
It should be emphasised that at ${\cal
O}(\lambda^2)$ one has
$\delta\lambda_0=\delta\lambda_2=-3\lambda^2 T_a^{(0)}/2,$ as
dictated by perturbation theory.  Also, one can identify in these
equations the contribution from each skeleton diagram separately.

The condition for cancellation of the overall
divergences in the equation of state 
(\ref{state-div}) determines $\delta m^2_0$ and $\delta\lambda_4$:
\bea
0&=&\delta m_0^2+\frac{1}{2}(\lambda+\delta\lambda_2)
\left[T_a^{(2)}+(m^2-M_0^2)T_a^{(0)}\right]+
\frac{\lambda^2}{6}S_{a,\textnormal{div}}(0)\nonumber\\
&&+(m^2-M_0^2)\left\{\frac{\lambda^2}{2}\left[
(T_a^{(0)})^2+T_a^{(I)}+\frac{1}{2}\tilde D_1\right]+
\frac{1}{4}\big(\lambda+\delta\lambda_2+\lambda^2T_a^{(0)}\big)D_0\right\},
\label{Eq:dm0}
\\
0&=&\frac{\delta\lambda_4}{3}+\frac{1}{2}\lambda
(\lambda+\delta\lambda_2)
\big(T_a^{(0)}+\lambda T_a^{(I)}\big)
+\frac{1}{2}\lambda^3\Big[(T_a^{(0)})^2+T_a^{(I)}+
\lambda\left(T_a^{(0)}T_a^{(I)}
+T_a^{(I,2)}\right)\Big]\nonumber\\
&&
+\frac{1}{4}\lambda\Big[\left(\lambda+\delta\lambda_2+\lambda^2T_a^{(0)}\right)
(D_0+\lambda D_1)+\lambda^2(\tilde D_1+\lambda D_2)\Big],
\label{Eq:dl4}
\eea

In addition, the subdivergence cancellation in the equation of state 
leads to a consistency check between $\delta\lambda_0$ and 
$\delta\lambda_2$. The simplest way to obtain it is as follows. Denoting by
$T_\textnormal{div}(T_F)$ and $S_{0,\textnormal{div}}(T_F)$ the contribution
from the tadpole and the setting-sun integrals proportional to $T_F[G],$
the subdivergence cancellation in (\ref{state-div}) reads as
\be
\frac{1}{2}(\lambda+\delta\lambda_2)T_\textnormal{div}(T_F)+
\frac{1}{2}\delta\lambda_2 T_F[G]+
\frac{1}{6}\lambda^2S_{0,\textnormal{div}}(T_F)=0.
\ee 
One should compare this with the condition for subdivergence cancellation
in the propagator equation (\ref{prop-div-a}):
\be
\frac{1}{2}(\lambda+\delta\lambda_0)T_\textnormal{div}(T_F)+
\frac{1}{2}\delta\lambda_0T_F[G]
+\frac{1}{6}\lambda^2S_{0,\textnormal{div}}(T_F)=0.
\ee
Using (\ref{Eq:Tad_div_BS}), from the comparison it follows that
\be
(\delta\lambda_0-\delta\lambda_2)\left[1+\frac{\lambda}{2}
\left(T_a^{(0)}+\frac{1}{2}D_0\right)\right]=0.
\label{Eq:consistency}
\ee
It is not obvious, when one looks at (\ref{dl2-a}) and (\ref{dl0-a}),
that this relation is 
fulfilled, since when one takes from these equations
the difference of the two quartic counter
couplings, one finds
\bea
\delta\lambda_0-\delta\lambda_2&=&
\frac{\lambda^3}{2}\left[1+\frac{\lambda}{2}
\left(T_a^{(0)}+\frac{1}{2}D_0\right)\right]^{-1}
\Bigg\{T_a^{(I)}+\frac{1}{2}D_1+\lambda\left(T_a^{(I,2)}+\frac{1}{2}D_2\right)
\nonumber
\\
&+&\frac{1}{2}\lambda^2\left[\left(T_a^{(I,2)}+\frac{1}{2}D_2\right)
\left(T_a^{(0)}+\frac{1}{2}D_0\right)-
\left(T_a^{(I)}+\frac{1}{2}D_1\right)
\left(T_a^{(I)}+\frac{1}{2}\tilde D_1\right)
\right]\Bigg\}.
\label{Eq:consistency2}
\eea
Comparing (\ref{Eq:consistency}) and (\ref{Eq:consistency2}) one has
\be
\delta\lambda_0=\delta\lambda_2,
\ee
and the requirement of the vanishing of the expression in the curly bracket 
of (\ref{Eq:consistency2}). This condition relates several
cut-off dependent integrals which can be (and should be) checked only
when the numerical solutions for $G(p)$ and for the
Bethe-Salpeter-like equation (\ref{Eq:BS}) are found.

With a fully analogous line of thinking one can investigate 
the consistency of the cancellation conditions of the divergences 
proportional with $v^0$ and $v^2$ in the propagator equation and 
the equation of state. Using (\ref{Eq:consistency}) the consistency
requirement leads to the relations:
\be
\delta m_2^2= \delta m_0^2,\qquad
\delta\lambda_4=3(\delta\lambda_2+\lambda^2T_a^{(0)}).
\ee
These two relations can be obtained also by comparing (\ref{dm2-a})
with (\ref{Eq:dm0}) and (\ref{dl2-a}) with (\ref{Eq:dl4}).
Since $\delta\lambda_0=-3 \lambda^2 T_a^{(0)}/2,$ 
to ${\cal O}(\lambda^2)$ all three quartic coupling counterterms are equal, 
as is expected from perturbative renormalisation.

\section{Analysis of the $O(N)$ model at the two-loop level of the
    truncation}

The broken symmetry phase of the $O(N)$ model represents a non-trivial
test-ground for checking the consistency of our counterterm determination.
It has two independent excitations: the $(N-1)$-fold degenerate ``pion''
multiplet and the more massive ``sigma''. Therefore there are four
renormalisability conditions for each of the two propagators. Since from the
counterterms introduced in Section~2 only five enter the two propagator
equations, one has to check if the remaining three conditions are fulfilled.
Although the algebra of this calculation is quite challenging with help of
symbolic computer calculations we could check that the extra conditions do
not impose any new relation on the counterterms, being identically
satisfied.

The equations for the inverse propagators $i G^{-1}_\sigma(p)=p^2-
\Sigma_0^\sigma(p)$ and $i G^{-1}_\pi(p)=p^2-\Sigma_0^\pi(p)$ 
lead to the following equations for the respective self-energies:
\bea
\nonumber
\Sigma_0^\pi(p)&=&m^2+\delta m_2^2+
(\lambda+\delta\lambda_2^A) \frac{v^2}{6N} + \frac{1}{6N}
\left[(N+1)\lambda+(N-1)\delta\lambda_0^A+2\delta\lambda_0^B\right] T_\pi
+\frac{1}{6N}\left(\lambda+\delta\lambda_0^A\right)T_\sigma\\
&&+\frac{\lambda^2 v^2}{9N^2}I_{\sigma\pi}(p),
\label{Eq:SP_0_ON}
\\
\Sigma_0^\sigma(p)&=&m^2+\delta m_2^2+
\left(3\lambda+\delta\lambda_2^A+2\delta\lambda_2^B\right)\frac{v^2}{6N} 
+\frac{1}{6N}\left(3\lambda+\delta\lambda_0^A+2\delta\lambda_0^B\right)T_\sigma
+\frac{N-1}{6N}\left(\lambda+\delta\lambda_0^A\right)T_\pi\nonumber\\
&&+\frac{\lambda^2 v^2}{18N^2}[9I_{\sigma\sigma}(p)+(N-1)I_{\pi\pi}(p)].
\label{Eq:SS_0_ON}
\eea
Here one ought to recall that the pion propagator determined 
self-consistently in the 2PI approximation does not obey 
Goldstone's theorem e.g. $\Sigma_0^\pi(p=0)\neq 0$. 
There exists a proposition at Hartree level \cite{ivanov05} to modify 
the 2PI functional to cure this deficiency, which one might 
attempt to generalise when including ${\cal O}(\lambda^2)$ 
skeleton diagrams. The method of counterterm resummation 
presented here would work equally well for such modified 2PI functionals.
 
When looking at (\ref{Eq:SS_0_ON}), one can make the following identifications of the respective
finite tadpole and bubble contributions:
\bea
\displaystyle
M_\pi^2&=&m^2+\frac{\lambda}{6N} v^2+
\frac{\lambda}{6N}\big[T_{\sigma, F}+(N+1)T_{\pi, F}\big],
\quad\ \ \ \Pi_0^\pi(p)=\frac{\lambda^2v^2}{9N^2}I_{\sigma\pi}^F(p),
\nonumber\\
\displaystyle
M_\sigma^2&=&m^2+\frac{\lambda}{2N}v^2+
\frac{\lambda}{6N}\big[3T_{\sigma, F}+(N-1)T_{\pi, F}\big],
\quad \ \Pi_0^\sigma(p)=\frac{\lambda^2v^2}{18N^2}
\big[9I_{\sigma\sigma}^F(p)+(N-1)I_{\pi\pi}^F(p)\big].\ \ 
\label{eq:MPI-sigpi}
\eea
The conditions for the renormalisability of the two self-energies 
$\Sigma_0^\pi(p)$ and $\Sigma_0^\sigma(p)$ read as
\bea
0&=&\delta m_2^2+\frac{\delta\lambda_2^A}{6N}v^2+
\frac{1}{6N}\left[(N+1)\lambda+(N-1)\delta\lambda_0^A+2\delta\lambda_0^B\right]
T_{\pi,\textnormal{div}}
+\frac{1}{6N}\left(\lambda+\delta\lambda_0^A\right)T_{\sigma,\textnormal{div}}
\nonumber\\
&&+\frac{1}{6N}\left[(N-1)\delta\lambda_0^A+2\delta\lambda_0^B\right] 
T_{\pi, F}
+\frac{\delta\lambda_0^A}{6N}T_{\sigma, F}
+\frac{\lambda^2 v^2}{9N^2}T_d^{(0)},
\label{Eq:G_p_rencon}
\\
0&=&\delta m_2^2+\left(\delta\lambda_2^A+2\delta\lambda_2^B\right)
\frac{v^2}{6N}+
\frac{1}{6N}\left(3\lambda+\delta\lambda_0^A+2\delta\lambda_0^B\right)
T_{\sigma,\textnormal{div}}
+\frac{N-1}{6N}(\lambda+\delta\lambda_0^A) T_{\pi,\textnormal{div}}
\nonumber
\\
&&+\frac{1}{6N}\left(\delta\lambda_0^A+2\delta\lambda_0^B\right)T_{\sigma,F}
+\frac{N-1}{6N} \delta\lambda_0^A T_{\pi, F}
+\frac{\lambda^2 v^2}{18N^2}(N+8)T_d^{(0)}.
\label{Eq:G_s_rencon}
\eea
For the three bubble integrals we used the analogue of
(\ref{bubble-renorm}) and the fact that their divergence
does not depend on the kind of fields which constitute them.

In analogy with the case of the one-component scalar field, one can
write down the explicit expressions for $T_{\pi,\textnormal{div}}$ and
$T_{\sigma,\textnormal{div}}$ expanding, as in (\ref{eq:tadpole-sing}), 
the exact propagators in the tadpole integrals around the common 
auxiliary propagator. One obtains
\bea
T_{\pi,\textnormal{div}}&=&T_d^{(2)}+(M_\pi^2-M_0^2)T_d^{(0)}+
\frac{\lambda^2v^2}{9N^2}T_{\pi,d}^{\sigma\pi(I)},
\label{Tp_div}
\\
T_{\sigma,\textnormal{div}}&=&T_d^{(2)}+(M_\sigma^2-M_0^2)T_d^{(0)}+
\frac{\lambda^2 v^2}{2N^2}T_{\sigma,d}^{\sigma\sigma(I)}+
\frac{\lambda^2 v^2}{18 N^2} (N-1) T_{\sigma,d}^{\pi\pi(I)}.
\label{Ts_div}
\eea
Although, the divergence involving integrands proportional to the finite
bubbles $I^F_{\alpha\beta}$ does not depend on the kind of fields which
constitute them, in the notation above we explicitly keep track of the
bubble which produces the divergence. The upper double index gives the two
propagators constituting the bubble, while lower index refers to the
self-energy they contribute to. This distinction will be exploited at the
end of this section to generalise, to the present case, the diagrammatic
interpretation of the consistency of the renormalisation conditions given at
the end of Section~3.

Substituting (\ref{Tp_div}) and (\ref{Ts_div}) into the
renormalisation conditions, one uses the defining equations of
$M_\sigma^2$ and $M_\pi^2$ and requires again separate vanishing of
the coefficients of the divergent pieces proportional to $v^0, v^2,
T_{\pi, F},$ and $T_{\sigma, F}$, respectively. For the pion
propagator one finds:
\bea
\label{eq:dm2}
0&=&\delta m_2^2+\frac{1}{6N}\left[(N+2)\lambda+N\delta\lambda_0^A+
2\delta\lambda_0^B\right] \left[T_d^{(2)}+(m^2-M_0^2)T_d^{(0)}\right],\\
0&=&\delta\lambda_2^A+
\left[(N+1)\lambda+(N-1)\delta\lambda_0^A+2\delta\lambda_0^B\right]
\left[\frac{\lambda}{6N} T_d^{(0)}+
\frac{\lambda^2}{9N^2} T_{\pi,d}^{\sigma,\pi(I)}\right]
\nonumber\\
&&
+(\lambda+\delta\lambda_0^A)\left[
\frac{\lambda}{2 N} T_d^{(0)}+
\frac{\lambda^2}{2N^2}T_{\sigma,d}^{\sigma\sigma(I)}+
\frac{\lambda^2}{18 N^2} (N-1) T_{\sigma,d}^{\pi\pi(I)}
\right]+\frac{2\lambda^2}{3 N} T_d^{(0)},
\label{eq:dl2A}
\\
0&=&(N-1)\delta\lambda_0^A+2\delta\lambda_0^B+
\frac{\lambda}{6N}\left\{(N-1)\left[(N+2)(\lambda+\delta\lambda_0^A)+
2(\lambda+\delta\lambda_0^B)\right]
+4(\lambda+\delta\lambda_0^B)\right\}T_d^{(0)},\ \ 
\label{eq:dl0B}
\\
0&=&\delta\lambda_0^A+\frac{\lambda}{6N}
\big[(N+2)(\lambda+\delta\lambda_0^A)+2(\lambda+\delta\lambda_0^B)\big]
T_d^{(0)}.
\label{eq:dl0A}
\eea
The last two equations form a closed set for $\delta\lambda_0^A$ and
$\delta\lambda_0^B$. Their expressions can be used in the first two
equations to obtain $\delta m^2_2$ and $\delta\lambda_2^A$. 

In case of the sigma propagator, the corresponding set of
renormalisation conditions reproduces exactly the same equation for
the mass-renormalisation $\delta m^2_2$. 
The condition for vanishing of the divergent coefficients 
of $T_{\pi, F}$ and $T_{\sigma, F}$ produces formally different expressions  
when compared to those in (\ref{eq:dl0B}) and (\ref{eq:dl0A}), but 
their solution for $\delta\lambda_0^A$ and $\delta\lambda_0^B$
coincides with the solution of 
(\ref{eq:dl0B}) and (\ref{eq:dl0A}). The only
 new condition arises from the vanishing of the coefficient of
$v^2,$ which reads as
\bea
0&=&\delta\lambda_2^A+2\delta\lambda_2^B+
\frac{\lambda}{6N}T_d^{(0)}\big[2(N+8)\lambda+(N+2)(\lambda+\delta\lambda_0^A)
+6(\lambda+\delta\lambda_0^B)\big]\nonumber\\
&&+\frac{\lambda^2}{18N^2}\left\{2(\lambda+\delta\lambda_0^B)\left[(N-1)
T_{\sigma, d}^{\pi\pi(I)}+9T_{\sigma, d}^{\sigma\sigma(I)}\right]
\right.\nonumber\\
&&\qquad\qquad\!\left.
+(\lambda+\delta\lambda_0^A)\left[(N-1)(T_{\sigma, d}^{\pi\pi(I)}+
2T_{\sigma, d}^{\sigma\pi(I)})+9T_{\sigma, d}^{\sigma\sigma(I)}\right]
\right\}.
\label{eq:dl2B}
\eea
Using the expression of $\delta\lambda_2^A$ obtained from (\ref{eq:dl2A})
the equation above determines $\delta\lambda_2^B$.
Note, that the expressions for $\delta\lambda_0^A,
\delta\lambda_0^B$ and $\delta m^2_2$ are identical with those
obtained in the 2PI-Hartree approximation in \cite{fejos08}. 
The extra term in the effective action, compared to the 
2PI-Hartree approximation  is responsible for the fact that 
$\delta\lambda_2^A\ne\delta\lambda_0^A$ and 
$\delta\lambda_2^B\ne\delta\lambda_0^B$.

In the broken symmetry phase the equation of state is given by
\bea
0&=&m^2+\delta m_0^2+\frac{1}{6N}(\lambda+\delta\lambda_4)v^2+
\frac{N-1}{6N}\left(\lambda+\delta\lambda_2^A\right)T_\pi+
\frac{1}{6N}\left(3\lambda+\delta\lambda_2^A+2\lambda_2^B\right)T_\sigma
\nonumber\\
&&+\frac{\lambda^2}{18N^2}\Big[3S_{\sigma\sigma\sigma}(0)+
(N-1)S_{\sigma\pi\pi}(0)\Big].
\label{Eq:EoS_ON}
\eea
The renormalisation conditions emerging from this equation naturally
separate into two groups. The vanishing of the divergent coefficients
proportional to $v^0$ and $v^2$ provides the expressions of
$\delta m_0^2$ and $\delta\lambda_4$. The consistent renormalisability
of the present truncation of the 2PI-approximation depends on the
vanishing of the subdivergences proportional to $T_{\pi, F}$ and
$T_{\sigma, F}$. There are no extra free parameters to ensure this.

In order to explicitly write the renormalisation conditions, one needs
the divergences of the two setting-sun integrals.
To obtain them, one does the same steps as in (\ref{eq:tadpole-sing}) 
and derives expressions similar to (\ref{deltaG}) for   
$\delta G_\sigma(p)=G_\sigma(p)-G_{P V}(p)$ and 
$\delta G_\pi(p)=G_\pi(p)-G_{P V}(p),$ namely: 
\bea
\delta G_\sigma(p)&=&i G_{P V}^2(p)\left[M_0^2-M_\sigma^2
-\frac{\lambda^2 v^2}{18 N^2}
\Big( 9I_{\sigma\sigma}^F(p)+(N-1) I_{\pi\pi}^F(p)\Big)
\right]
+G_\sigma^r(p),
\label{Eq:delta_Gs}
\\
\delta G_\pi(p)&=&i G_{P V}^2(p)\left(M_0^2-M_\pi^2
-\frac{\lambda^2 v^2}{9 N^2} I_{\sigma\pi}^F(p)
\right)+G_\pi^r(p).
\label{Eq:delta_Gp}
\eea
The regular parts $G_\sigma^r(p)$ and $G_\pi^r(p)$ behave
asymptotically as $p^{-6}.$ Using
$G_\alpha(p)=G_{P V}(p)+\delta G_\alpha(p)$ with
$\alpha=\{\sigma,\pi\},$ one finds the following expressions for the 
two types of setting-sun integrals in which the parts contributing to the
divergences are well separated:
\bea
S_{\sigma\sigma\sigma}(0)&=&S_{P V}(0)+3\int_k\delta G_\sigma(k)
I_{P V}^{\sigma\sigma}(k)+S_{\sigma\sigma\sigma}^{F(1)},\nonumber\\
S_{\sigma\pi\pi}(0)&=&S_{P V}(0)+\int_k\delta G_\sigma(k) I_{P V}^{\pi\pi}(k)+
2\int_k \delta G_\pi(k)I_{P V}^{\sigma\pi}(k)+S_{\sigma\pi\pi}^{F(1)}.
\label{Eq:sss_spp}
\eea
For the bubble $I_{PV}(k)$ we indicated the original
two propagators which were replaced by auxiliary ones. 

When substituting (\ref{Eq:delta_Gs}) and (\ref{Eq:delta_Gp}) 
into (\ref{Eq:sss_spp}) one replaces $I^F_{\alpha\beta}(p)$ with
$I_{PV,F}^{\alpha\beta}(p)$ ($\alpha\in\{\pi,\sigma\}$), 
since the difference gives a finite contribution. 
The same steps as performed on (\ref{Eq:SS_decomp})
give for the divergent parts of the setting-sun diagrams:
\bea
S_{\sigma\sigma\sigma}^\textnormal{div}(0)&=&
S_{P V}(0)+3T_d^{(0)}\left(T_{\sigma,F}-T_d^{(2)}\right)+3(M_\sigma^2-M_0^2)
T_{\sigma,d}^{\sigma\sigma (I)}+\frac{\lambda^2 v^2}{6N^2}(N+8)T_d^{(I,2)},
\label{Eq:SSS_div}
\\
\nonumber
S_{\sigma\pi\pi}^\textnormal{div}(0)&=&
S_{P V}(0)+T_d^{(0)}\left(T_{\sigma,F}+2 T_{\pi,F}-3 T_d^{(2)}\right)
+(M_\sigma^2-M_0^2) T_{\sigma,d}^{\pi\pi (I)}
+2(M_\pi^2-M_0^2) T_{\pi,d}^{\sigma\pi (I)}\\
&&+\frac{\lambda^2 v^2}{18 N^2}(N+12) T_d^{(I,2)}.
\label{Eq:SPP_div}
\eea
The divergent expressions
$T_{\sigma,d}^{\sigma\sigma(I)},$ $T_{\sigma,d}^{\pi\pi(I)},$ and
$T_{\pi,d}^{\sigma\pi(I)}$ are the same as those occurring in (\ref{Tp_div}) and
(\ref{Ts_div}). For the divergent part of the integral
containing two bubbles we used the definition (\ref{eq:divI2})
without distinguishing which bubble produces it, because this is an 
overall divergence and does not show up in any consistency condition.

The condition for cancellation of the divergences in the equation of
state (\ref{Eq:EoS_ON}) reads as
\bea
\nonumber
0&=&\delta m_0^2+\frac{\delta\lambda_4}{6N}v^2+
\frac{N-1}{6N}(\lambda+\delta\lambda_2^A)T_{\pi, \textnormal{div}}+
\frac{N-1}{6N}\delta\lambda_2^A T_{\pi, F}\\
&&+(3\lambda+\delta\lambda_2^A+2\delta\lambda_2^B)
\frac{T_{\sigma,\textnormal{div}}}{6N}+
(\delta\lambda_2^A+2\delta\lambda_2^B)\frac{T_{\sigma, F}}{6N}
+\frac{\lambda^2}{18N^2}\Big[
3S_{\sigma\sigma\sigma}^\textnormal{div}(0)
+(N-1)S_{\sigma\pi\pi}^\textnormal{div}(0)\Big].
\label{Eq:EoS_rencond_ON}
\eea
The renormalisation conditions can be read from this equation 
using the expressions given in (\ref{Tp_div}), (\ref{Ts_div}), 
(\ref{Eq:SSS_div}), and (\ref{Eq:SPP_div}) with the corresponding 
formulae for $M_\sigma^2$ and $M_\pi^2$ given in (\ref{eq:MPI-sigpi}). 
The counterterms $\delta m_0^2$ and $\delta\lambda_4$ are determined
from the vanishing of the coefficients of $v^0$ and $v^2$, respectively.
The conditions for the cancellation of the coefficients of $T_{\sigma,F}$
and $T_{\pi,F}$ are given by
\bea
\nonumber
0&=&\frac{1}{6N}(\delta\lambda_2^A+2\delta\lambda_2^B)+
\frac{\lambda T_d^{(0)}}{36 N^2} \Big[3\lambda (N+8)+(N+2)
\delta\lambda_2^A+6\delta\lambda_2^B\Big]+
\frac{\lambda^3 (T_d^{(0)})^2}{108 N^3}(5N+22)\\
&&+\frac{\lambda^3}{N^3}\left(
\frac{1}{4}T_{\sigma,d}^{\sigma\sigma (I)}
+\frac{N-1}{36} T_{\sigma,d}^{\pi\pi (I)}
+\frac{N-1}{54}T_{\pi,d}^{\sigma\pi (I)}
\right),\\
\nonumber
0&=&\frac{\delta\lambda_2^A}{6N}+
\frac{\lambda T_d^{(0)}}{36 N^2} \Big[\lambda (N+8)+(N+2)
\delta\lambda_2^A+2\delta\lambda_2^B\Big]+
\frac{\lambda^3 (T_d^{(0)})^2}{108 N^3}(3N+10)\\
&&+\frac{\lambda^3}{6N^3}\left(
\frac{1}{2}T_{\sigma,d}^{\sigma\sigma (I)}
+\frac{N-1}{18} T_{\sigma,d}^{\pi\pi (I)}
+\frac{N+1}{9}T_{\pi,d}^{\sigma\pi (I)}
\right).
\eea

\begin{figure}[t]
\centerline{
\includegraphics[keepaspectratio,width=0.95\textwidth,angle=0]{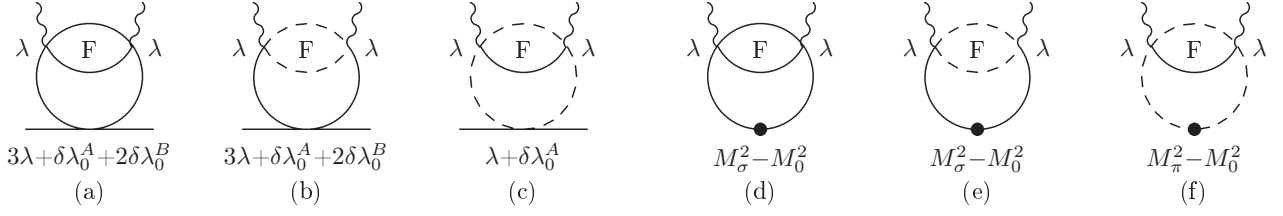}}
\caption{
Illustration of the appearance of the same radiative structures
producing the divergences denoted by
$T_{\sigma,d}^{\sigma\sigma(I)},$ $T_{\sigma,d}^{\pi\pi(I)},$ and
$T_{\pi,d}^{\sigma\pi(I)}$ in the equation for the sigma
propagator and the subdivergences of the equation of state. Solid
(dashed) lines indicate that the auxiliary propagator $G_{PV}$ comes
from a (pseudo)scalar propagator. For other notations we refer to
Fig.~\ref{Fig:1comp}.}
\label{Fig:Ncomp}
\end{figure}

Using the solution for $\delta\lambda_2^A$ and $\delta\lambda_2^B,$ as
obtained from (\ref{eq:dl2A}) and (\ref{eq:dl2B}), and the solution for
$\delta\lambda_0^A$ and $\delta\lambda_0^B$ obtained from
(\ref{eq:dl0B}) and (\ref{eq:dl0A}), one can easily verify that the
above two consistency conditions are fulfilled.  Note, that the
coefficients of $T_{\sigma,d}^{\sigma\sigma (I)},$
$T_{\sigma,d}^{\pi\pi (I)},$ and $T_{\pi,d}^{\sigma\pi (I)}$
separately vanish when the solution for the coupling counterterms are
used. The reason is the same as the one given at the end of
Section~3, namely the appearance of the same structure in the
renormalisation conditions of both the propagator equations and of the
equation of state. The diagrammatic illustration given in
Fig.~\ref{Fig:Ncomp} parallels that of Fig.~\ref{Fig:1comp}.  Diagrams
(a), (b) and diagram (c) correspond to the last two terms of
(\ref{Ts_div}) and the last term of (\ref{Tp_div}), respectively,
while diagram (d) and diagrams (e) and (f) correspond to the last
but one term in (\ref{Eq:SSS_div}) and the last two terms of the first
line of (\ref{Eq:SPP_div}), respectively.

\section{Analysis of the $O(N)$ model with ``basket-ball'' diagrams included}

At the ${\cal O}(\lambda^2)$ truncation level of the effective action the 
finite self-energies are decomposed as
\bea
\Sigma_\pi(p)=M_\pi^2+\Pi_0^\pi(p)+\Pi_2^\pi(p),
\qquad
&&\Pi_2^\pi(p)=\frac{\lambda^2}{18 N^2} \left[
S^F_{\sigma\sigma\pi}(p)+(N+1) S^F_{\pi\pi\pi}(p)\right],\\
\Sigma_\sigma(p)=M_\sigma^2+\Pi_0^\sigma(p)+\Pi_2^\sigma(p),
\qquad
&&\Pi_2^\sigma(p)=\frac{\lambda^2}{18 N^2}
\left[3 S^F_{\sigma\sigma\sigma}(p)+(N-1) S^F_{\sigma\pi\pi}(p)\right],
\eea
where $M_\pi^2,$ $M_\sigma^2,$ $\Pi_0^\pi(p)$ and $\Pi_0^\sigma(p)$ 
are defined in (\ref{eq:MPI-sigpi}). Since the infrared sector has no
influence on the asymptotic behaviour of the self-energies, one can
separate a common leading asymptotics in the part of the self-energies
containing the contribution of the setting-sun integrals 
($\alpha\in\{\pi,\sigma\}$):
\be
\Pi_2^\alpha(p)=\Pi_a(p)+\Pi_{2,0}^\alpha(p)+\Pi_r^\alpha(p), 
\qquad \lim_{p\rightarrow\infty}
\frac{\Pi_a(p)}{p^2}\sim (\ln p)^\beta,\qquad \lim_{p\rightarrow \infty}
\frac{\Pi^\alpha_{2,0}(p)}{p^2}\rightarrow 0,\qquad 
\Pi_r^\alpha(p)\sim p^{-2}.
\ee
The corresponding decomposition of the finite contribution of the setting-sun  
integrals reads as
\bea
S_{\alpha\beta\beta}^F(p)=S_{a,F}(p)+
S_{\alpha\beta\beta}^{0,F}(p)+S_{\alpha\beta\beta}^r(p),
\eea
where $\alpha,\beta\in\{\pi,\sigma\}.$

Compared to (\ref{Eq:G_p_rencon}) and (\ref{Eq:G_s_rencon})
the renormalisation conditions for the equations of
 $G_\pi^{-1}(p)$ and  $G_\sigma^{-1}(p)$ 
are now completed, respectively, by the following terms: 
\be
-p^2 \delta Z + \Sigma_\pi^\textnormal{div}(p),\qquad
-p^2 \delta Z + \Sigma_\sigma^\textnormal{div}(p),
\label{Eq:suplement}
\ee
where
\be
\Sigma_\pi^\textnormal{div}(p)=\frac{\lambda^2}{18 N^2} 
\left[ S^\textnormal{div}_{\sigma\sigma\pi}(p)+
(N+1) S^\textnormal{div}_{\pi\pi\pi}(p)\right],\qquad
\Sigma_\sigma^\textnormal{div}(p)=
\frac{\lambda^2}{18 N^2}
\left[3 S^\textnormal{div}_{\sigma\sigma\sigma}(p)+
(N-1) S^\textnormal{div}_{\sigma\pi\pi}(p)\right].
\ee
The renormalisation condition (\ref{Eq:EoS_rencond_ON}) for the 
equation of state remains unchanged. Evidently, in
(\ref{Eq:G_p_rencon}) and (\ref{Eq:G_s_rencon}) one has to replace 
$T_d^{(0)}$ by $T_a^{(0)}.$

With the same steps which lead from (\ref{Eq:azonossag_G}) to 
(\ref{Eq:azonossag2_G}) one can write the following representation for
the propagators
\be
G_\alpha(p)=G_a(p)+\delta G_\alpha(p),\quad
\delta G_\alpha(p):=
-i G_a^2(p)\left[M_\alpha^2-M_0^2+\Pi_{0}^\alpha(p)+
\Pi_{2,0}^\alpha(p)\right]+G_r^\alpha(p),
\label{Eq:azonossag_G_ON}
\ee
where $G_r^\alpha(p)$ behaves asymptotically as $p^{-6}.$

Using (\ref{Eq:azonossag_G_ON}), the divergences of the tadpole
integrals can be easily separated
\be
\begin{split}
T_\sigma&=T_a^{(2)}+(M^2_\sigma-M_0^2)T_a^{(0)}+
\frac{\lambda^2 v^2}{18N^2} (N+8) T_a^{(I)}-
i\int_k G_a^2(k)\Pi_{2,0}^\sigma(k)\Big|_\textnormal{div}+T_{\sigma,F},
\\
T_\pi&=T_a^{(2)}+(M^2_\pi-M_0^2)T_a^{(0)}+
\frac{\lambda^2 v^2}{9N^2} T_a^{(I)}-
i\int_k G_a^2(k)\Pi_{2,0}^\pi(k)\Big|_\textnormal{div}+T_{\pi,F}.
\end{split}
\label{Eq:Tad-SP}
\ee

The same procedure which led to (\ref{eq:S0F-finite}) provides now
 an integral equation for the finite piece of the
setting-sun integrals $S_{\alpha\beta\beta}^{0,F}(p)$ which grows no
faster than logarithmically for asymptotic values of $p$. Combining
the corresponding contributions of different setting-sun integrals one
obtains the following set of coupled integral representations:
\be
\begin{split}
\Pi_{2,0}^\sigma(p)&=-\frac{i}{18 N^2}\int_k G_a^2(k) K(p,k)
\left[
F_\sigma(k)+A_{\sigma\sigma} \Pi_{2,0}^\sigma(k) + 
A_{\sigma\pi}\Pi_{2,0}^\pi(k)\right], \\
 \Pi_{2,0}^\pi(p)&=-\frac{i}{18 N^2}\int_k G_a^2(k) K(p,k)
\left[
F_\pi(k)+A_{\pi\sigma} \Pi_{2,0}^\sigma(k) + A_{\pi\pi} \Pi_{2,0}^\pi(k)
\right].
\end{split}
\label{Eq:ON_system}
\ee
$K(p,k)$ was defined in (\ref{Eq:BS_kernel}). Here we introduced the
matrix $A$ with elements 
$A_{\sigma\sigma}=N+8,$ $A_{\sigma\pi}=2(N-1),$ 
$A_{\pi\sigma}=2,$ $A_{\pi\pi}=3N+4,$ and also the two
inhomogeneous terms of the above relations:
\be
\begin{split}
F_\sigma(k)&=A_{\sigma\sigma}(M_\sigma^2-M_0^2)+A_{\sigma\pi}(M_\pi^2-M_0^2)+
\frac{\lambda^2 v^2}{18 N^2} B_\sigma I_{a,F}(k),\\
F_\pi(k)&=A_{\pi\sigma}(M_\pi^2-M_0^2)+A_{\pi\pi}(M_\pi^2-M_0^2)+
\frac{\lambda^2 v^2}{18 N^2} B_\pi I_{a,F}(k),
\end{split}
\ee
with $B_\sigma=(N+8)^2+4(N-1)$ and $B_\pi=8(N+3)$.

The solution of (\ref{Eq:ON_system}) takes the form
\be
\Pi_{2,0}^\alpha(p)=-\frac{i}{18N^2} \int_k G_a^2(k)
\Gamma_{\alpha\beta}(p,k) \left[A_{\beta\gamma} (M^2_\gamma-M_0^2 u_\gamma)
+\frac{\lambda^2}{18 N^2}B_\beta I_{a,F}(k)\right],
\label{Eq:Pi20_sol}
\ee
where $\alpha,\beta,\gamma\in\{\pi,\sigma\},$ $u_\pi=u_\sigma=1.$
$\Gamma_{\alpha\beta}(p,k)$ satisfies the following
Bethe-Salpeter-like matrix equation
\be
\Gamma_{\alpha\beta}(p,k)=K(p,k)\delta_{\alpha\beta}
-\frac{i}{18N^2} \int_q K(p,q) A_{\alpha\gamma} \Gamma_{\gamma\beta}(q,k).
\ee

One observes that $\Gamma_{\sigma\sigma}(p,k)$ couples only to 
$\Gamma_{\pi\sigma}(p,k)$ and $\Gamma_{\pi\pi}(p,k)$ couples to 
$\Gamma_{\sigma\pi}(p,k).$ Exploiting that 
$A_{\sigma\sigma}-A_{\pi\sigma}=A_{\pi\pi}-A_{\sigma\pi},$
one can verify that the two combinations 
$Q_1(p,k)=\Gamma_{\sigma\sigma}(p,k)-\Gamma_{\pi\sigma}(p,k)$ and 
$Q_2(p,k)=\Gamma_{\pi\pi}(p,k)-\Gamma_{\sigma\pi}(p,k)$ satisfy the same
equation
\be
Q_i(p,k)=K(p,k)-c\frac{i}{18 N^2}\int_q K(p,q) Q_i(q,k),
\label{Eq:Q12}
\ee
where $c=N+6.$ This means that
$Q_1(p,k)=Q_2(p,k),$ and consequently
there is a relation between the elements of the $\Gamma(p,k)$ matrix:
\be
\Gamma_{\sigma\sigma}(p,k)-\Gamma_{\pi\sigma}(p,k)=
\Gamma_{\pi\pi}(p,k)-\Gamma_{\sigma\pi}(p,k).
\label{Eq:Gamma_rel1}
\ee
Another relation which can be derived reads as
\be
\Gamma_{\sigma\sigma}(p,k)+(N-1)\Gamma_{\pi\sigma}(p,k)=
\Gamma_{\pi\pi}(p,k)+(N-1)^{-1}\Gamma_{\sigma\pi}(p,k).
\label{Eq:Gamma_rel2a}
\ee
This is because the combinations on the left- and right-hand sides of
the equation both satisfy an equation similar to (\ref{Eq:Q12}), 
but with $c$ replaced by $\tilde c=3(N+2).$ 
Making use of  (\ref{Eq:Gamma_rel1}) in (\ref{Eq:Gamma_rel2a}) one obtains
\be
\Gamma_{\sigma\pi}(p,k)=(N-1) \Gamma_{\pi \sigma}(p,k).
\label{Eq:Gamma_rel2}
\ee
We shall see, that the relations (\ref{Eq:Gamma_rel1}) and 
(\ref{Eq:Gamma_rel2}) play an important role in checking some of the 
consistency relations of our renormalisation procedure.

Having obtained the integral representation (\ref{Eq:Pi20_sol}) for 
$\Pi_{2,0}^\sigma(p)$ and $\Pi_{2,0}^\pi(p),$ one uses them in 
(\ref{Eq:Tad-SP}) to write the divergent part of the tadpole integrals
with help of the new divergent matrices $D_i^{\alpha\gamma}$ to be defined 
below:
\be
T_{\alpha,\textnormal{div}}=T_a^{(2)} u_\alpha +
\left( T_a^{(0)} \delta_{\alpha\gamma}+D_0^{\alpha\gamma} 
A_{\beta\gamma}\right) (M_\gamma^2-M_0^2 u_\gamma)
+\frac{\lambda^2 v^2}{18 N^2}\left( T_a^{(I)} A_{\alpha\sigma}+
D_1^{\alpha\beta} B_\beta\right).
\label{Eq:Tdiv_matrix}
\ee
The same new divergent quantities appear in the divergent
part of the self-energies containing the contributions
of the setting-sun integrals. One obtains with the procedure which gave
(\ref{Eq:Sdiv_final}) the following expression:
\bea
\nonumber
\Sigma_\alpha^\textnormal{div}(p)&=&
(N+2)\frac{\lambda^2}{18 N^2} S_a(p) u_\alpha
+\left(\frac{\lambda^2 v}{18 N^2}\right)^2 \left[
\left(T_a^{(0)} T_a^{(I)}+ T_a^{(I,2)} \right) \delta_{\alpha\gamma}+
A_{\alpha\beta} \left(T_a^{(0)}D_1^{\beta\gamma}+D_2^{\beta\gamma}\right)
\right] B_\gamma
\\
&+&\frac{\lambda^2}{18 N^2} A_{\alpha\beta} 
\left\{
\left[
\Big(\big(T_a^{(0)}\big)^2+T_a^{(I)}\Big)\delta_{\beta\rho}
+\left(T_a^{(0)} D_0^{\beta\gamma}+\tilde D_1^{\beta\gamma}\right)
A_{\gamma\rho}\right](M_\rho^2-M^0 u_\rho)
+T_a^{(0)} T_{\beta,F}
\right\}.
\label{Eq:Sdiv_matrix}
\eea
The divergent quantities $D_i^{\alpha\beta}$ with 
$i\in\{0,1,2\}$ and $\tilde D_1^{\alpha\beta}$ are obvious
generalisations of those appearing
 in (\ref{Eq:D01_def}) and (\ref{Eq:D1T2_def})  
with $\Gamma_0^{\alpha\beta}$ and $\Gamma_1^{\alpha\beta}$ given by
\[
\Gamma_0^{\alpha\beta}(p)=-\frac{i}{18 N^2}\int_k
\Gamma_{\alpha\beta}(p,k)G_a^2(k),\qquad 
\Gamma_1^{\alpha\beta}(p)=-\frac{i}{18 N^2}\int_k 
\Gamma_{\alpha\beta}(p,k)G_a^2(k)I_{a,F}(k).
\]
These matrices inherit from $\Gamma_{\alpha\beta}$
some relations between its elements which are analogous to
(\ref{Eq:Gamma_rel1})
and (\ref{Eq:Gamma_rel2}):
\begin{alignat}{3}
&D_i^{\sigma\sigma}-D_i^{\pi\sigma}=D_i^{\pi\pi}-D_i^{\sigma\pi},\qquad
&&\tilde D_1^{\sigma\sigma}-\tilde D_1^{\pi\sigma}=
\tilde D_1^{\pi\pi}-\tilde D_1^{\sigma\pi},
\label{Eq:rel1}
\\
&D_i^{\sigma\pi}=(N-1) D_i^{\pi\sigma},\qquad
&&\tilde D_1^{\sigma\pi}=(N-1) \tilde D_1^{\pi\sigma}.
\label{Eq:rel2}
\end{alignat}

In order to determine the counterterms one has to use
(\ref{Eq:Tdiv_matrix}) and (\ref{Eq:Sdiv_matrix}) in the
renormalisation conditions for the two propagators
(\ref{Eq:G_p_rencon}), (\ref{Eq:G_s_rencon}), supplemented with the
terms given in (\ref{Eq:suplement}), and in the renormalisation
condition of the equation of state (\ref{Eq:EoS_rencond_ON}). In what
follows we review, without giving the lengthy explicit expressions, 
how one can obtain the counterterms and the consistency conditions
similar to (\ref{Eq:consistency2}).

The vanishing of the coefficients of 
$v^0,$ $v^2,$ $T_{\sigma,F},$
and $T_{\pi,F}$ in the two renormalisation conditions for $G_\pi(p),$ 
$G_\sigma(p)$ results in eight conditions for the determination of five 
counterterms: $\delta m_2^2,$ $\delta\lambda_0^A,$ $\delta\lambda_0^B,$
$\delta\lambda_2^A,$ and $\delta\lambda_2^B.$ Out of them 
$\delta\lambda_2^A$ and $\delta\lambda_2^B$ are obtained from the
 vanishing of the coefficients of $v^2$. 
The vanishing of the divergent coefficients of $T_{\sigma,F}$
and $T_{\pi,F}$ in any one of the propagators
determines $\delta\lambda_0^A$ and $\delta\lambda_0^B.$
The condition to obtain the same expressions from  the
renormalisation conditions of the other propagator reduces to the equation 
\bea
\nonumber
\lambda^2 (N+2)\left[
C_1 T_a^{(0)} - C_0 T_a^{(I)}
-3 (N+2)\left(C_0 D_1^{\pi\pi}-C_1 D_0^{\pi\pi} + 
D_1^{\sigma\pi} D_0^{\pi\sigma}-D_0^{\sigma\pi} D_1^{\pi\sigma}\right)
\right]\\
+6 N (\lambda C_1 + N C_0)=0,
\eea
where $C_i=(N-1) D_i^{\pi\sigma}-D_i^{\sigma\pi}.$ Using (\ref{Eq:rel2})
this consistency condition is identically satisfied. The same way, in order
to obtain the same expression for $\delta m_2^2$ 
from the vanishing of the coefficients of $v^0$
in either one of the two renormalisation conditions for the
propagators one needs 
\bea
\nonumber
\left[
\lambda^2 (N+6) T_a^{(0)}+ 6 N \left(\lambda+\delta\lambda_0^B\right)
\right]
\left(D_0^{\sigma\sigma}+D_0^{\sigma\pi}-D_0^{\pi\sigma}-D_0^{\pi\pi}
\right)\\
+\lambda^2 (N+6) \left(
\tilde D_1^{\sigma\sigma}+\tilde D_1^{\sigma\pi}-
\tilde D_1^{\pi\sigma}-\tilde D_1^{\pi\pi}
\right)=0.
\eea
By (\ref{Eq:rel1}), this is again identically satisfied.  As a partial
summary we emphasise that the eight conditions renormalising the
propagator equations of the $O(N)$ model in the broken phase are
algebraically satisfied with the five counterterms, independent of the
form of the asymptotic propagator $G_a$.

Turning now to the equation of state, one can subtract its
renormalisation condition from the renormalisation condition of 
$G_\sigma,$ and require the cancellation of the term proportional to
$T_{\sigma,F}, T_{\pi,F},$ $v^0$ and $v^2.$ 
We can easily find the following relations:
\be
\delta\lambda_2^A=\delta\lambda_0^A, \ \ 
\delta\lambda_2^B=\delta\lambda_0^B, \qquad
\delta m_0^2= \delta m_2^2,\qquad
\delta\lambda_4=\delta\lambda_2^A+2\delta\lambda_2^B+
\frac{\lambda^2}{3N}(N+8) T_a^{(0)}.
\ee
The first two conditions above impose two consistency conditions on the
coupling counterterms if one uses their expression as obtained from the
renormalisation conditions for the propagators. These are complicated
expressions, similar to that in (\ref{Eq:consistency2}), which now depend
also on $N$ and relate several cut-off dependent integrals. As in the
one-component case, these relations can be checked only numerically.

\section{Conclusions}

In this paper we have extended the renormalisation procedure developed in
Ref.~\cite{fejos08} to the case of momentum-dependent self-energy appearing
when one goes beyond the Hartree-Fock level of truncation of the 2PI
effective action. Two types of momentum-dependent truncations of scalar
$\Phi^4$ type models were investigated. In the first case the asymptotic
behaviour of the exact propagator remains unchanged relative to the
tree-level, while in the second case the asymptotic behaviour of the exact
propagator changes.

Working in the broken symmetry phase we introduced infrared safe auxiliary
propagators having the right asymptotic behaviour in order to explore the
divergence structure of the model. Beyond the overall divergences
proportional to the zeroth and second power of the background field we found
only such subdivergences which were proportional to the finite part of the
tadpole integrals. This is the reason that no counterterms are needed
beyond those appearing at the Hartree truncation of the 2PI effective
potential. Requiring the vanishing of the coefficients of independent
divergences, we gave explicit expressions for the counter couplings of the
model in terms of some integrals of the auxiliary propagator.

For the first type of truncation investigated, our method reproduces the
expressions of the counter couplings which were previously reported in the
literature for the case of a real selfinteracting scalar field, but with no
reference to any Bethe-Salpeter equation. Although, the divergence structure
was known in the literature for the second type of truncation, the explicit
expressions for the counter couplings, which are needed to make the
equations of the theory finite, appear to our best knowledge for the first
time in this paper. These expressions depend in a non-trivial way on the
solution of a Bethe-Salpeter-like equation and can be determined only
numerically. This was beyond the scope of our present study in which we
concentrated mainly on the structural aspects of the renormalisation.

The proposed method treats the conditions for overall and subdivergence
cancellation on equal footing in the propagator equation(s) and the equation
of state. Therefore, the number of conditions is larger than the actual
number of counterterms. As shown in this paper, these conditions are
algebraically fulfilled for the first type of truncation, both in the case
of a one-component and a multi-component scalar field (O(N) model). For the
second type of truncation at least some part of the consistency conditions
can be verified only numerically. It is quite impressive, though, that in
this case the consistency of the propagator renormalisation can be proved
algebraically. In the explicit demonstration the internal structure of the
Bethe-Salpeter-like matrix equation plays an essential role. It is the
compatibility of the renormalised propagator equations with the renormalised
equation of state which leads to those consistency equations which can be
verified only numerically.

The auxiliary propagators applied for the separation of divergences
introduce also the renormalisation scale $M_0$. The explicit dependence of
the counter couplings on this scale allows to establish the renormalised
trajectory of the 2PI approximation in a higher-dimensional coupling space
corresponding to the larger number of independent bare couplings necessary
for its consistent renormalisation than in the perturbative case. Even when
one restricts the beta-functions to leading (e.g. ${\cal O}(\lambda^2)$)
order one finds the perturbative beta-function only for the coefficient of
the $v^4$ term in the 2PI-functional. It is interesting to note that beyond
the perturbatively higher order contributions to the counter couplings also
the appearance of the asymptotic propagator $G_a(p)$ in the relevant
integrals produces deviations from the perturbative behaviour of the
beta-functions determined in the 2PI approximation.

When passing to the numerical investigations the spirit of our paper
suggests to make use of the explicit expressions for the cut-off dependent
counterterms in the cut-off regulated equations (for instance
(\ref{prop-eq}) and (\ref{state-eq})) and verify numerically the cut-off
independence of their solution. Alternatively, one can prescribe values for
a number of physical quantities in terms of the renormalised one- and
two-point functions. One attempts then the determination of the asymptotic
behaviour of the propagator from a finite twice-subtracted dispersive
representation satisfying the renormalisation prescriptions. Before
proceeding to the solution of the self-consistent equations in the
non-asymptotic regime one has to check the consistency relations between
asymptotic integrals. Numerical investigation of these questions essential
for the validity of the renormalisation methods proposed for the 2PI
approximation, will be reported elsewhere.

\appendix

\section{Appendix A:
Divergence structures for the two-loop truncation of the 
2PI approxi\-ma\-tion}

The divergences of the integrals defined in (\ref{Eq:tadpole}),
(\ref{Eq:bubble}) and (\ref{Eq:setting-sun}) can be most conveniently 
explored using Pauli-Villars type terms obtained by adding and
subtracting in the integrands the auxiliary propagator 
\be   
i G_{P V}^{-1}(p)=p^2-M_0^2.   
\label{eq:aux-prop}
\ee 

The simplest is to separate the logarithmic divergent part of the 
bubble integral from its finite part. They read as
\be   
I_\textnormal{div}=-i\int_k G_{P V}^2(k)\Big|_\textnormal{div}=:T_d^{(0)},
\qquad    
I_F(p)=-i\int_k\big[G(k)G(k+p)-G_{P V}^2(k)\big],   
\label{bubble-renorm}   
\ee   
where the divergent piece is clearly independent of the external 
momentum. 

For the propagator appearing in (\ref{eq:G}) one uses the identity
\be
G(p)=G_{PV}(p)-iG(p)G_{PV}(p)   
\left(M^2-M_0^2+\frac{\lambda^2}{2}v^2I_F(p)\right).     
\label{eq:tadpole-sing} 
\ee
Solving (\ref{eq:tadpole-sing}) for $G(p)$ and expanding
it around $G_{P V}(p)$ this equality can be rearranged also as   
\be   
G(p)=G_{P V}(p)+\delta G(p),\qquad
\delta G(p)=-i G_{P V}^2(p)\left(M^2-M_0^2+\frac{\lambda^2v^2}{2}I_F(p)\right)+
G_r(p), 
\label{deltaG}  
\ee   
where $G_r(p)$ is the regular part of the
propagator, which contains at least three powers of $G_{P V}(p)$.
It behaves asymptotically as $\sim p^{-6}$ and produces a convergent 
contribution when integrated over momentum.

With help of the representation (\ref{deltaG}) the divergences and 
the finite part of the tadpole integral are easily separable.
After replacing $I_F(p)$ by $I_{P V,F}(p)$
in (\ref{deltaG})  (since the contribution of the difference 
to the tadpole is finite) the following partitioning is found:
\be
T[G]=T[G_{P V}]+(M^2-M_0^2)T_d^{(0)}-\frac{1}{2}\lambda^2v^2i\int_p
G_{P V}^2(p)I_{P V,F}(p)+T_F[G].
\ee
Here one introduces the notations
\be
T[G_{P V}]=\int_p G_{P V}(p)\Big|_\textnormal{div}=: T_d^{(2)},\qquad 
T_d^{(I)}:=-i\int_p G_{P V}^2(p)I_{P V,F}(p)\Big|_\textnormal{div},
\label{eq:Tp_div}
\ee
and obtains the first entry in (\ref{TIS-div}).

The setting-sun integral is rewritten after the replacement $G(p)=G_{P V}(p)
+\delta G(p)$ is made for each of the three propagators as 
\be
S(p=0,G)=S_{P V}(0)+3\int_k\delta G(k)I_{P V}(k)+S_F^{(1)},
\label{Eq:SS_decomp}
\ee
where 
$S_{PV}(0)=-i\int_k\int_qG_{PV}(k)G_{PV}(q)G_{PV}(k+q)\Big|_\textnormal{div}.$
In the integral above one decomposes $I_{P V}(k)$ into a divergent and
a finite part. For the piece proportional to the divergent part $I_{P
V,\textnormal{div}}=T_d^{(0)}$ one uses $\delta G(k)=G(p)-G_{P V}(k),$
while when taking the finite part of the bubble one substitutes
$\delta G(k)$ from (\ref{deltaG}). With help of (\ref{eq:Tp_div}) the
following representation for the divergent part is obtained:
\be
S_\textnormal{div}(0,G)=S_{P V}(0)+3(T[G]-T_d^{(2)})T_d^{(0)}+
3(M^2-M_0^2) T_d^{(I)} -\frac{3}{2}\lambda^2 v^2 
i\int_k G^2_{P V}(k) I_F(k) I_{P V,F}(k)\Big|_\textnormal{div}.
\ee
The divergent part of the integral above can be obtained by
replacing $I_F(k)$ by $I_{P V,F}(k)$ (the difference gives a finite 
contribution). Then, introducing another cut-off dependent integral 
\be
T_d^{(I,2)}:=-i\int_k G_{P V}^2(k)I^2_{P V,F}(k)\Big|_\textnormal{div},
\label{eq:divI2}
\ee
one obtains the last entry in (\ref{TIS-div}).

\section{Divergence structures for the truncation of the 2PI approximation
at ${\cal O}(\lambda^2)$  skeleton diagram level}
 
The analysis goes quite in parallel with that given in Appendix A.
One defines the basic divergent integrals to be used in the
exploration of the divergence structure:
\begin{alignat}{2}
\nonumber
&T_a^{(2)}:=\int_p G_a(p)\Big|_\textnormal{div}, &
\qquad  &T_a^{(0)}:=-i\int_p G_a^2(p)\Big|_\textnormal{div},\\
&T_a^{(I)}:=-i\int_p G_a^2(p)I_{a,F}(p)\Big|_\textnormal{div}, &\qquad 
&T_a^{(I,2)}:=-i\int_p G_a^2(p)I_{a,F}^2(p)\Big|_\textnormal{div}.
\end{alignat}

Making use of (\ref{eq:G-a}), (\ref{Eq:Pi2_decomp}) and (\ref{Eq:Ga_def}),
the identity corresponding to (\ref{eq:tadpole-sing}) reads
\be
G(p)=G_a(p)-i G(p)G_a(p)\left[M^2-M_0^2+\Pi_0(p)+\Pi_{2,0}(p) 
+\Pi_r(p)\right].
\label{Eq:azonossag_G}
\ee
Solving (\ref{Eq:azonossag_G}) for $G(p)$ and expanding
around $G_a(p)$ one obtains
\be
G(p)=G_a(p)+\delta G(p),\quad
\delta G(p):=
-i G_a^2(p)\left[M^2-M_0^2+\frac{1}{2}\lambda^2 v^2 I_F(p)+\Pi_{2,0}(p)\right]+G_r(p),
\label{Eq:azonossag2_G}
\ee   
where $G_r(p)$ contains also the contribution of $\Pi_r(p).$
Power counting shows that for asymptotically large values of $p$ one has 
$G_r(p)\sim p^{-6}.$

Eq.~(\ref{Eq:azonossag2_G}) is used to separate the divergence of
the tadpole integral (see (\ref{tadpole-a})), also to analyse the
divergence structure of the momentum-dependent setting-sun integral
and, eventually 
to obtain the integral representation of $S_{0,F}(p)$ (e.g. $\Pi_{2,0}(p)$).

\section*{Acknowledgements}   
Work supported by the Hungarian Scientific Research Fund (OTKA) under   
Contract Nos. T046129 and T068108. Zs. Sz. is supported by OTKA Postdoctoral   
Grant no. PD 050015.

}%allowysplaybreaks ends here
\end{document}